\documentclass[a4paper,12pt]{article}

\usepackage[usenames]{color}
\usepackage{graphicx}
\usepackage{amsmath}
\usepackage{amssymb}
\usepackage{setspace}

\begin{document}
\begin{titlepage}
\null
\begin{flushright}
May, 2012
\end{flushright}

\vskip 1.8cm
\begin{center}
 
  {\Large \bf Sigma Model BPS Lumps on Torus\\ 
}

\vskip 1.8cm
\normalsize

  {\bf Atsushi Nakamula\footnote{nakamula(at)sci.kitasato-u.ac.jp} 
and Shin Sasaki\footnote{shin-s(at)kitasato-u.ac.jp}}

\vskip 0.5cm

  { \it 
  Department of Physics \\
  Kitasato University \\
  Sagamihara 252-0373, Japan
  }

\vskip 2cm

\begin{abstract}
We study doubly periodic Bogomol'nyi-Prasad-Sommerfield (BPS) lumps in supersymmetric $\mathbb{C}P^{N-1}$ 
non-linear sigma models on a torus $T^2$. 
Following the 
philosophy of the Harrington-Shepard construction of calorons in Yang-Mills
 theory, we obtain the $n$-lump solutions on compact spaces by suitably
 arranging the $n$-lumps on $\mathbb{R}^2$ at equal intervals.
We examine the modular invariance of the solutions and find that there are no
 modular invariant solutions for $n=1, 2$ in this construction.
\end{abstract}
\end{center}
\end{titlepage}

\newpage

\section{Introduction}
Instantons in Yang-Mills theories at finite temperature have been
extensively investigated in the past years. Instantons at finite
temperature, commonly called calorons, were firstly studied by Harrington
and Shepard \cite{HaSh},
who demonstrated the analytic description to
the 1-instanton of $SU(2)$ gauge theory living in $\mathbb{R}^3 \times S^1$. The
radius of the compact space $S^1$ is naturally interpreted as the inverse of
the temperature $T$.
The Harrington-Shepard caloron is constructed by locating the infinitely
many BPST instantons \cite{BePoScTy} along 
the one direction with the equal separation $T^{-1}$.
The authors of \cite{HaSh}  start from the BPST instanton in the 't Hooft ansatz:
\begin{equation}
A^c_m = \eta^c_{mn} \partial^n \log \phi (x), \quad \phi (x) = 1 +
 \frac{\rho^2}{(x_m - a_m)^2},
\label{eq:HS_calorons}
\end{equation}
where $c=1,2,3$ is the index of $su(2)$, 
$\eta^c_{mn}$ is the 't Hooft symbol and 
$a_m$, $\rho$ are the position and the size of the instanton, respectively.
It can easily be shown that the anti-self-dual equation is enjoyed
if the function $\phi(x)$ obeys the Laplace equation.
Hence the 1-instanton in $\mathbb{R}^3 \times S^1$ is obtained by 
superposing the BPST instantons placed periodically 
along the $x^4$ direction with period $1/T$.
By scaling the size of each instanton $\rho\to\rho/\sqrt{2\pi T}$, 
one can perform the infinite sum in $\phi (x)$ as 
\begin{equation}
\phi (\vec{x}, T) = 1 +\frac{\rho^2}{2\pi T} \sum_{k = - \infty}^{\infty} 
 \frac{1}{\vec{x}^2 + (x_4 - k T^{-1})}=
1+\frac{\rho^2}{2r}\frac{\sinh(2\pi Tr)}{\cosh(2\pi Tr)-\cos(2\pi Tx_4)},
\end{equation}
where $\vec{x} = (x_1,x_2,x_3)$, $r=\sqrt{\vec{x}^2}$ and 
we have taken $a_m = 0$ for simplicity.
Therefore calorons are interpreted as the periodic instantons 
on $\mathbb{R}^4$ \cite{Wi}.
However, the Harrington-Shepard construction cannot be applied to the 
general solutions including all the moduli parameters.
This is because the 't Hooft ansatz does
not contain all the moduli. 
To find the most general solutions, 
one needs to consider the Nahm construction \cite{Na} of calorons, which provides a strong
scheme to study the structure of solutions or moduli spaces.
A natural generalization of calorons are doubly periodic instantons on a
torus $T^2$. Instantons on a torus, sometimes called torons, are studied in various
contexts \cite{Ba, Ja}.

Sometimes problems in gauge theories are 
simplified when one considers non-linear sigma models 
that are recognized as the strong gauge coupling limit of the UV
theories. Actually, explicit constructions of instantons or calorons are possible
in non-linear sigma models.
For example, instantons of the sigma models in two dimensions have very
simple structures. These two-dimensional instantons are called lumps.
The lumps in the sigma models are studied in much detail \cite{Le, Wa}, 
where the explicit construction of lumps, moduli space structure, and
scattering process  has been investigated.
Recently, the constituent structure of the lumps in the non-linear sigma models
is studied \cite{Br, BrBrJaWiWo, CoTo}.
In \cite{Br}, it is discussed that 
the lumps with twisted boundary conditions 
in compact spaces lead to the constituent
structure.
This type of structure of the lumps on $\mathbb{R} \times S^1$ is quite
similar to the calorons in Yang-Mills theories with nontrivial
holonomies, in which there appear monopole constituents of calorons \cite{Calorons}.

The aim of this paper is to establish the systematic construction of BPS lumps in
supersymmetric non-linear sigma models on a torus $T^2$ for the arbitrary
charge $n$. The sigma model lumps with $n \ge 2$ on $\mathbb{R}^2$
are obtained by multiplying the charge-1 solution by $n$ times. 
We will show that the same is true even for the lumps in compact spaces.
Following the Harrington-Shepard philosophy, we will collect 
the lumps on $\mathbb{R}^2$
aligned in two distinct directions and 
construct the explicit solutions on $T^2$. 
We also demonstrate that the collections of infinitely many $n$ lumps on $\mathbb{R}^2$ 
with various boundary conditions result in the solutions on $T^2$ with 
twisted periodic conditions.
The solutions have appropriate pole structure and 
correct topological charge $n$.
We will also examine the modular invariance of the solutions.

The organization of this paper is as follows. 
In section 2, we define the model. We consider the supersymmetric
$\mathbb{C}P^{N-1}$ model and the BPS equation for lumps.
In section 3, focusing on the $\mathbb{C}P^1$ model, 
we give the constructive method to fabricate doubly periodic BPS lumps on a torus 
starting from the charge-$n$ ones on $\mathbb{R}^2$. 
The modular invariance of the solutions will be studied. 
Section 4 is devoted to the conclusion and
discussions.

\section{$\mathbb{C}P^{N-1}$ sigma model and BPS equations}
In this section, we start from the 
$\mathcal{N} = 1$ supersymmetric 
$\mathbb{C}P^{N-1} \sim SU(N)/[SU(N-1)\times U(1)]$ 
sigma model in four dimensions.
Although supersymmetry is not essential for the construction of
solutions, we embed the model into the superfield formalism.
This is because one can easily generalize the model to the ones with
other target spaces in the superspace formalism \cite{HiNi}. 
Another important point is that the supersymmetric property of solutions 
is necessary when one discusses the relations between the sigma model
lumps and other solitonic objects in gauge theories. 
See footnote \ref{fn:susy}.
We follow the Wess-Bagger conventions \cite{WeBa}. 
The space-time metric is given by $\eta_{mn} = \mathrm{diag}(-1,+1,+1,+1)$. 
Following the quotient construction of sigma-models \cite{HiNi}, 
the Lagrangian in four-dimensional $\mathcal{N} = 1$ superspace is given by
\begin{eqnarray}
\mathcal{L} = \int \! d^4 \theta 
\left(
\Phi_i^{\dagger} e^{2V} \Phi_i - c V
\right), \quad (i=1, \cdots, N), 
\label{eq:Lag_sup}
\end{eqnarray}
where the chiral superfields $\vec{\Phi} = \Phi_i$ are the fundamental
representation $(\mathbf{N})$ of the global $SU(N)$ symmetry, 
$V$ is the $U(1)$ vector superfield and 
$c > 0$ is the Fayet-Iliopoulos (FI) parameter. 
The component expansion of the chiral superfield is given by 
\begin{eqnarray}
\Phi_i (y, \theta) = \phi_i (y) + \sqrt{2} \theta \psi_i (y) + \theta^2 F_i (y), 
\end{eqnarray}
while the the vector superfield in the Wess-Zumino gauge is 
\begin{eqnarray}
V = - \theta \sigma^m \bar{\theta} A_m + i \theta \theta
 \bar{\theta} \bar{\lambda} - i \bar{\theta} \bar{\theta} \theta \lambda
 + \frac{1}{2} \theta^2 \bar{\theta}^2 D.
\end{eqnarray}
In the following, we consider the bosonic part of the Lagrangian \eqref{eq:Lag_sup}.
The Lagrangian in the component form is given by 
\begin{eqnarray}
\mathcal{L} &=& - (D_m \phi_i) (D^m \phi_i)^{\dagger} 
+ D (\phi_i \bar{\phi}_i - c) + F_i \bar{F}_i,
\end{eqnarray}
where $D_m * = \partial_m * + i A_m *$ is the $U(1)$ gauge covariant derivative.
From the D-term condition, we have the constraint for the the scalar fields,
\begin{eqnarray}
|\phi_i|^2 = c,
\label{constraint}
\end{eqnarray}
while the F-term condition is trivial. Therefore the Lagrangian is
rewritten as 
\begin{eqnarray}
\mathcal{L} = - |D_m \phi_i|^2, \quad  |\phi_i|^2 = c.
\end{eqnarray}
Since the gauge field does not have the kinetic term, it is eliminated by
the equation of motion,
\begin{eqnarray}
A_m &=& i \frac{c^{-1}}{2} (\bar{\phi}_i \partial_m \phi_i - \partial_m \bar{\phi}_i
 \phi_i).
\end{eqnarray}
Next, we consider the BPS equation for lumps 
which depends on the two-dimensional directions $x^a \ (a = 1,2)$. 
The lumps are instantons in two-dimensional sigma models.
In the following, we consider two-dimensional models though we started
from four dimensions. 
The dimensional reduction from four to two dimensions is straightforward.
The model becomes the two-dimensional $\mathcal{N} = (2,2)$
$\mathbb{C}P^{N-1}$ model\footnote{
\label{fn:susy}
This $\mathcal{N} = (2,2)$ $\mathbb{CP}^{N-1}$ sigma model 
is the world-volume effective theory of a 1/2 BPS vortex in
supersymmetric gauge theory in four dimensions \cite{HaTo}.
It was discussed in \cite{EtIsNiOhSa} that 
the 1/2 BPS lumps in the supersymmetric sigma model are 
interpreted as the 1/4 BPS instantons in supersymmetric gauge theories.
Clearly, this lumps/instantons correspondence is based on the
supersymmetric setup.
}.
The energy is given by 
\begin{eqnarray}
E &=& \int \! d^2 x  \ 
\left[ \frac{1}{2} \left| D_a \phi_i \pm i 
\varepsilon_{ab} D_b \phi_i  \right|^2 \pm i \varepsilon_{ab} \left( D_a 
\phi_i \right) \left( D_b \phi_i \right)^{\dagger} \right] \nonumber \\
&\ge& \pm \int \! d^2 x i \varepsilon_{ab} \left( D_a \phi_i \right) 
\left( D_b \phi_i \right)^{\dagger} \nonumber \\
&=& \pm 2 \pi c Q
\label{eq:energy_bound}
\end{eqnarray}
where $\varepsilon_{12} = -1$ is the antisymmetric epsilon symbol 
and the topological charge $Q$ has been
defined as  
\begin{eqnarray}
Q &=& \frac{1}{2 \pi c} \int \! d^2 z \ \left( |D_z \phi_i |^2 - 
|D_{\bar{z}} \phi_i |^2 \right)
\nonumber \\
&=& - \frac{1}{4 \pi} \int \! d^2 x \ \varepsilon^{ab} F_{ab}.
\end{eqnarray}
Here the complex coordinate in two dimensions is defined as 
$z = \frac{1}{\sqrt{2}} (x^1 + i x^2)$. 
The gauge field and the covariant derivative 
are complexified in the same way.
From the energy bound in eq.~\eqref{eq:energy_bound}, the BPS condition is
given by 
\begin{eqnarray}
D_a \phi_i \pm i \varepsilon_{ab} D_b \phi_i = 0, \label{eq:BPS}
\end{eqnarray}
or equivalently,
\begin{eqnarray}
 D_{\bar{z}} \phi = 0, \quad D_{z} \phi = 0.
\label{eq:BPSc}
\end{eqnarray}
The first and the second conditions 
correspond to the plus and minus
signs in eq.~\eqref{eq:BPS} respectively.
In the following, we focus on the first condition.
The solutions to the BPS equation \eqref{eq:BPSc} preserve a half of
$\mathcal{N} = (2,2)$ supersymmetry. Therefore the lumps are 1/2 BPS configurations.

In order to satisfy the constraint \eqref{constraint}, 
it is convenient to consider the following field decomposition:
\begin{eqnarray}
\phi_i = W_i \frac{\sqrt{c}}{ \sqrt{W^{\dagger}_j W_j}},
\end{eqnarray}
where $W_i$ is an $N$-component vector.
Then one easily finds that the BPS equation becomes
\begin{eqnarray}
D_{\bar{z}} \phi_i = \sqrt{c} P (\partial_{\bar{z}} W_i) (W^{\dagger} \cdot
 W)^{-1/2} = 0,
\label{eq:wBPSeq}
\end{eqnarray}
where $P_{ij} \equiv 1_{ij} - W_i \frac{1}{W^{\dagger} \cdot W}
W^{\dagger}_j$ is the projection operator. 
Therefore solutions to the BPS equation 
are given by holomorphic functions $W_i = W_i (z)$ \cite{MaSu}.
Using the gauge symmetry, we fix the gauge as 
\begin{eqnarray}
W_i = 
\left(
\begin{array}{c}
1 \\
w_{\hat{i}}
\end{array}
\right), \quad (\hat{i}=2, \cdots, N).
\end{eqnarray}
The topological charge for the BPS lump is, therefore, given by 
\begin{eqnarray}
Q &=& \frac{1}{2 \pi c} \int \! d^2 z \ 
\frac{c}{W^{\dagger} \cdot W} \partial_{\bar{z}} W^{\dagger} P
\partial_z W
\nonumber \\
&=&  \frac{1}{2 \pi} \int \! d^2 z \ 
\frac{1}{(1 + |w_{\hat{i}}|^2)^2}
\left[
(1 + |w_{\hat{i}}|^2) |\partial w_{\hat{i}}|^2 - w_{\hat{i}} \bar{\partial} \bar{w}_{\hat{i}} 
\bar{w}_{\hat{j}} \partial w_{\hat{j}}
\right].
\end{eqnarray}
Since we have the relation $\partial \bar{\partial} \log (W^{\dagger} \cdot W) =
\frac{1}{W^{\dagger} \cdot W} \bar{\partial} W^{\dagger}_i P_{ij}
\partial W_j$, the topological charge is rewritten as 
\begin{eqnarray}
Q &=& \frac{1}{2 \pi} \int \! d^2 z \ 
\partial \bar{\partial} \log (W^{\dagger} \cdot W)
\nonumber \\
&=& \frac{1}{4 \pi} 
 \int \! d^2 z \ 
\left[
\partial \bar{\partial} \log (W^{\dagger} \cdot W)
+ 
\bar{\partial} \partial \log (W^{\dagger} \cdot W)
\right]
\nonumber \\
&=& \frac{i}{4\pi} \oint \! 
\left[
\bar{\partial} \log (W^{\dagger} \cdot W)  d\bar{z}
- 
\partial \log (W^{\dagger} \cdot W)  d z
\right].
\end{eqnarray}
Therefore the topological charge is determined by the residue of 
the function $U \equiv \partial \log (W^{\dagger} \cdot W)$,
\begin{equation}
Q = \frac{1}{2} 
\left(
\mathrm{Res}_z (U) + \mathrm{Res}_{\bar{z}} (\bar{U})
\right).
\end{equation}
Since the energy \eqref{eq:energy_bound} is invariant under the conformal transformation in 
the two-dimensional plane $\mathbb{R}^2$, the field is defined on the conformally
compactified $S^2$. The lumps are, therefore, harmonic maps from $S^2$ to
$\mathbb{C}P^{N-1}$ that are classified by integers, namely, the
topological charges.

\section{BPS lumps}
In this section, we give the constructive procedure to formulate the BPS lumps on a torus $T^2$
with appropriate base point conditions.
Before going to the totally compactified space $T^2$, we establish the 
relations between the lump solutions in $\mathbb{R}^2$ and $\mathbb{R} \times S^1$. 
In the following, we consider the $N=2$ case, namely, the $\mathbb{C}P^1$ model.
In this case, only the nontrivial component in $W_i$ is $w_2 \equiv u (z)$, and
the topological charge is given by 
\begin{equation}
Q = \frac{1}{2 \pi} \int \! d^2 z \
\frac{|\partial u|^2}{(1 + |u|^2)^2}.
\end{equation}

\subsection{Lumps on $\mathbb{R}^2$}
Let us start from the 1-lump solution 
on $\mathbb{R}^2$
denoted as $u^{(1)}$.
The solution to the BPS equation \eqref{eq:wBPSeq}
should be a holomorphic
function, and it is required to be settled down to the vacuum
asymptotically. 
When we take the base point (vacuum) condition $u^{(1)} (\infty) = 0$, 
the 1-lump solution is given by \cite{MaSu}
\begin{equation}
u^{(1)} (z) = \frac{\lambda}{z - \check{z}_1}, \quad \lambda \in \mathbb{R}, \ \check{z}_1 \in
 \mathbb{C}.
\label{eq:1-lump0}
\end{equation}
The residue of the function $U$ associated with the solution
\eqref{eq:1-lump0} is evaluated at the pole $z = \check{z}_1$, giving the
expected result $Q=1$.
When one considers a different base point condition, for example
$u^{(1)}(\infty) = 1$, the 1-lump solution is given by 
\begin{eqnarray}
u^{(1)} (z) = \frac{z - \hat{z}_1}{z - \check{z}_1}, \quad 
\hat{z}_1 \not= \check{z}_1, \quad 
\hat{z}_1, \check{z}_1 \in \mathbb{C}.
\label{eq:1-lump}
\end{eqnarray}
For this solution, the topological charge density $q$ is 
\begin{eqnarray}
q = \frac{1}{2\pi} 
\frac{\lambda^2}{(|z-z_1|^2 + \lambda^2)^2}, 
\end{eqnarray}
where we have defined the parameters 
$z_1 \equiv \frac{\hat{z}_1 - \check{z}_1}{2}$, $\lambda \equiv
\frac{|\hat{z}_1 - \check{z}_1|}{2}$, 
interpreted as the position
and the size of the lump. The profile of the energy density is found in
fig \ref{fig:lumps}.
Similarly, for the base point condition $u^{(1)} (\infty) = \infty$, we have the
1-lump solution $u^{(1)} (z) = \lambda (z - \hat{z}_1)$.

Generalizations to the multilump solutions are straightforward.
The $n$-lump solutions $u^{(n)}$ are obtained by multiplying the 1-lump 
solutions $n$ times. The result is meromorphic rational functions with
degree $n$.
For example in the case of the base point condition $u^{(n)} (\infty) = 1$, the solutions are given by
\begin{eqnarray}
u^{(n)} (z) = \prod^n_{k=1}
\frac{z - \hat{z}_k}{z - \check{z}_k}, \quad 
\hat{z}_k \not= \check{z}_j \ (\textrm{for any } j, k).
\label{N-lump-sol}
\end{eqnarray}
One can easily find that the residue of the function $U$ for the
solution \eqref{N-lump-sol} is $n$, which gives the desired result $Q=n$.

\subsection{Lumps on $\mathbb{R} \times S^1$}
Next, we consider the lumps on $\mathbb{R} \times S^1$ by 
compactifying one space-time direction.
Without loss of generality, one can consider the imaginary direction in
the complex plane $\mathbb{C}$ as the compact direction.
We expect that solutions on $\mathbb{R} \times S^1$ are interpreted as
periodically aligned lumps on $\mathbb{R}^2$.
Following the Harrington-Shepard philosophy, 
we multiply the infinite number of the 1-lump solutions \eqref{eq:1-lump0} 
located at the equal interval $\beta \in \mathbb{R}$ along the 
imaginary direction. Namely, we consider the following solution
\begin{eqnarray}
u^{(1)} (z, \beta) &=& \prod_{k=-\infty}^{\infty} \frac{\lambda}{z - z_0 - i
 \beta k} 
\nonumber \\
&=& \frac{\lambda}{z - z_0} \prod_{k=1}^{\infty}
 \frac{\lambda^2/\beta^2}{(z-z_0)^2/\beta^2 + k^2},
\end{eqnarray}
where we have multiplied by the 1-lump solutions so that the solution
$u^{(1)} (z, \beta)$ has poles at $z = z_0 + i \beta k$.
Since the infinite product of the 1-lump solution diverges, 
we employ the $\zeta$-function regularization to find the finite solution:
\begin{eqnarray}
\prod_{k=1}^{\infty} \frac{\lambda^2}{\beta^2} =  
(\lambda/\beta)^{-1/2}.
\label{eq:S1-reg}
\end{eqnarray}
After the regularization, we find 
that the solution on $\mathbb{R} \times S^1$ is obtained as
\begin{equation}
u^{(1)} (z, \beta) = 
\frac{1}{2} \frac{1}{\sinh \pi \beta^{-1} (z-z_0)}.
\label{eq:S1-1lump_reg}
\end{equation}
For this solution, the poles of the function $U$ are at $z = z_0$, and
it is easy to find that the topological charge for this solution is $Q = 1$.
Since the solution \eqref{eq:S1-1lump_reg} satisfies the antiperiodic
boundary condition $u^{(1)} (z + i \beta, \beta) = - u^{(1)} (z, \beta)$, the
solution is allowed only when the twisted boundary condition is
imposed.
This solution has been discussed in \cite{Br, BrBrJaWiWo} in the context of the constituent
structure of sigma model lumps on the compact space.
In \cite{Br, BrBrJaWiWo}, the authors introduced nontrivial holonomy
parameters in the solution \eqref{eq:S1-1lump_reg} and studied its 
partonic, or constituent, nature.

 \begin{figure}[t]
 \begin{center}
 \includegraphics[scale=0.6]{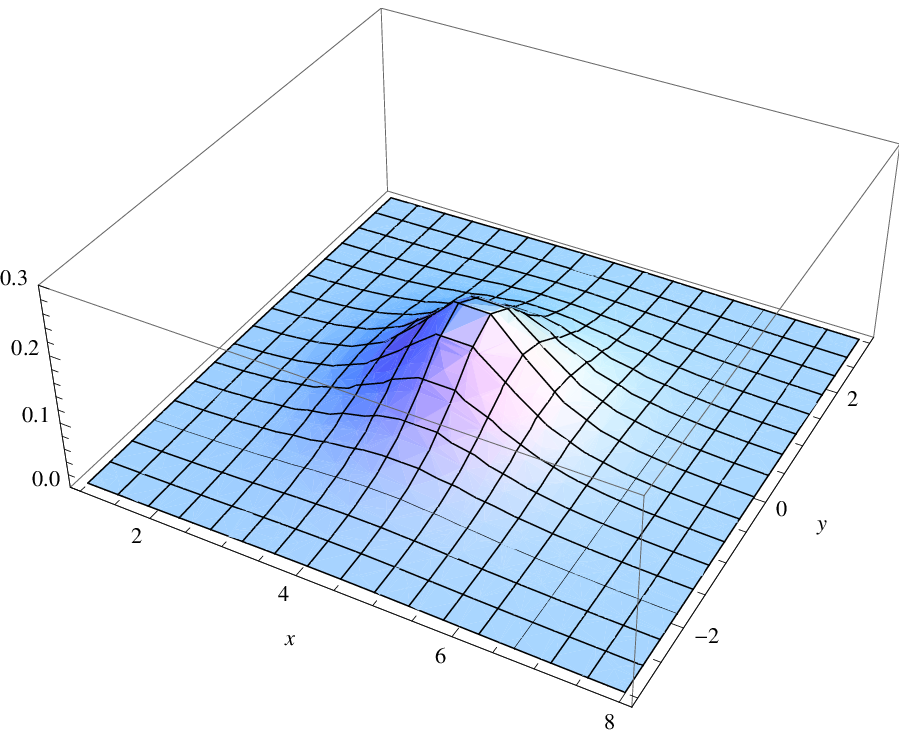} 
 \includegraphics[scale=0.6]{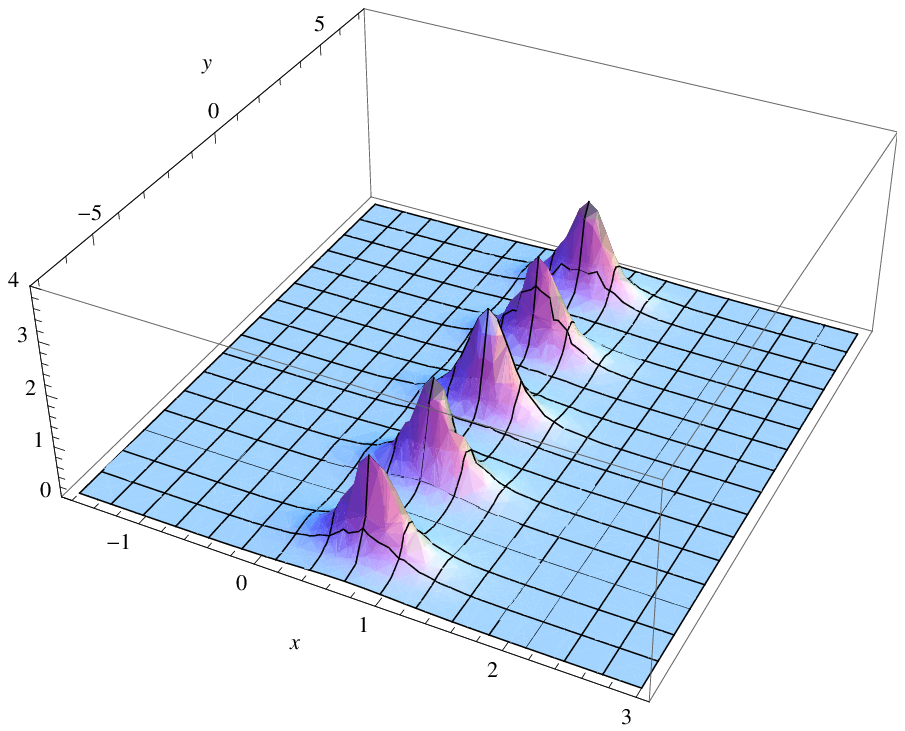}
 \includegraphics[scale=0.6]{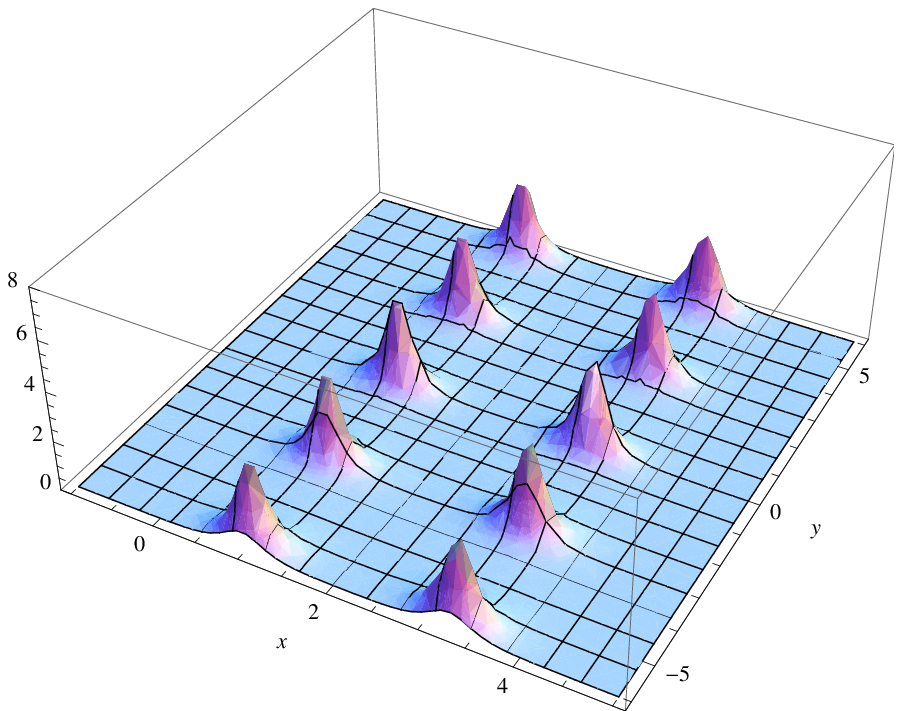}
 \end{center}
 \caption{
Charge density for the solution \eqref{eq:1-lump} with $\hat{z}_1
  = 4$, $\check{z}_1 = 2$ (left), for the solution \eqref{eq:S1-1lump} 
with $\beta = 2, \lambda = 0.4, \nu = 0.1$, (middle). 
Charge density for the solution \eqref{eq:S1-nlump}, $n=2$ case (right).
$\beta = 2, \lambda_1 = 0.4, \lambda_2 = 1.4, \nu_1 = 0.1, \nu_2 =  1.1$.}
 \label{fig:lumps}
 \end{figure}

Next, we consider the 1-lump solution \eqref{eq:1-lump} by choosing the base
point condition $u^{(1)}(\infty) = 1$. 
Again, we arrange the solution along $\check{z}_k = 2 \nu + i \beta k$,
$\beta \in \mathbb{R}$, $\lambda, \nu \in \mathbb{C}$, $k \in \mathbb{Z}$.
We further demand that the zeros 
of the solution appear at $\hat{z}_k = 2 \lambda + i
\beta k$. By choosing these zero points,  the size of the 1-lump solution 
\eqref{eq:1-lump} does not diverge at $k \to \infty$ 
and is fixed to be $|\lambda - \nu|$.
The position of the lump in one period is $\lambda - \nu$. 
Then we obtain the solution as follows: 
\begin{eqnarray}
u^{(1)} (z, \beta) &=& \prod_{k=-\infty}^{\infty} 
\left(
\frac{z - 2 \lambda - i \beta k}{z - 2 \nu - i \beta k}
\right)
\nonumber \\
&=& \frac{z - 2 \lambda}{z - 2 \nu} 
\prod_{k=1}^{\infty} 
\frac{k^2}{\beta^{-2} (z-2\nu)^2 + k^2} \frac{\beta^{-2} (z - 2 \lambda)^2 +
k^2}{k^2}
\nonumber \\
&=& \frac{\sinh \pi \beta^{-1} (z - 2 \lambda)}{\sinh \pi \beta^{-1} (z
 - 2 \nu)}.
\label{eq:S1-1lump}
\end{eqnarray}
Thanks to the good base point condition, we do not need any regularization.
Moreover, the solution preserves the periodic boundary condition,
\begin{eqnarray}
u^{(1)}
(z + i \beta, \beta) 
= 
u^{(1)} (z, \beta).
\end{eqnarray}
This solution was found in \cite{MoWi} in the same way we have just
shown above.
The energy profile for this solution is given in fig \ref{fig:lumps}.
Since the poles of the function $U$ are at $z = 2 \nu$ inside the one period, 
the residue is evaluated as $\mathrm{Res}_{z = 2 \nu} \partial
\log W^{\dagger} W = 1$ at the pole. 
Therefore the topological charge is given by $Q=1$.
One can easily find that the decompactification limit $\beta \to \infty$
of the solution gives the correct result:
\begin{eqnarray}
\lim_{\beta \to \infty} u^{(1)}(z, \beta) = \frac{z - 2 \lambda}{z - 2 \nu}.
\end{eqnarray}
The $n$-lump generalization is straightforward. 
This is obtained from the solution \eqref{N-lump-sol} on $\mathbb{R}^2$. 
The result is 
\begin{eqnarray}
u^{(n)} (z,\beta) = \prod_{k=1}^n
\frac{\sinh \pi \beta^{-1} (z - 2 \lambda_k)}
{\sinh \pi \beta^{-1} (z - 2 \nu_k)}.
\label{eq:S1-nlump}
\end{eqnarray}
The charge density profile for this solution 
is found in fig \ref{fig:lumps} for the $n=2$ case.

\subsection{Lumps on $T^2$}
In this subsection, we construct the multilump solutions 
on a torus $T^2$ 
by extending the superposition procedure established in the previous subsection.
By suitably arranging the solutions on $\mathbb{R} \times S^1$, we will
find the lump solutions with topological charges $Q=n \ge 1$. 
It is known that there is no harmonic map from 
the genus $g$ Riemann surface to $\mathbb{C}P^1 \sim S^2$ when the
degree $n$ of the map is less than $g$ \cite{Su, Kn, Eells}. 
Therefore we expect that there is no $Q=1$ periodic solution on a torus.
Actually, as we will see, the $n=1$ lump constructed below does not show
the doubly periodic property. 
When $n \ge 2$, the solutions can be doubly periodic and are rewritten
as elliptic functions. 
Even more, for the cases $n \ge 3$, the solutions show the modular invariance.

Let us begin with the solution \eqref{eq:S1-1lump} 
on $\mathbb{R} \times S^1$, a lump aligned in the
imaginary direction. In order to find solutions on $T^2$, 
we locate the solution \eqref{eq:S1-1lump} along the real direction
at the interval $\gamma$.
Assuming that $\gamma/\beta > 0$, the array of $\mathbb{R} \times S^1$ lumps with interval
$\gamma$ is given by 
\begin{eqnarray}
u^{(1)}(z, \beta, \gamma) &=& 
\prod_{k=- \infty}^{\infty} 
\frac{\sinh \pi \beta^{-1} (z - 2 \lambda - \gamma k)}{\sinh \pi
\beta^{-1} (z - 2 \nu - \gamma k)} 
\nonumber \\
&=& \frac{\sinh \pi \beta^{-1} (z - 2 \lambda)}{\sinh \pi \beta^{-1}
 (z-2 \nu)} 
\prod_{k=1}^{\infty}
\frac{
(1 - e^{- 2 \pi \beta^{-1}\gamma k} e^{2 \pi \beta^{-1} (z - 2 \lambda)})
(1 - e^{ 2 \pi \beta^{-1}\gamma k} e^{- 2 \pi \beta^{-1} (z - 2 \lambda)})
}
{
(1 - e^{- 2 \pi \beta^{-1}\gamma k} e^{2 \pi \beta^{-1} (z - 2 \nu)})
(1 - e^{ 2 \pi \beta^{-1}\gamma k} e^{- 2 \pi \beta^{-1} (z - 2 \nu)})
}.
\nonumber \\
\label{T2products}
\end{eqnarray}
We can rewrite this infinite product as the pseudo periodic
$\theta$ functions by using the formula
\footnote{
Here, $\mathrm{Im} z > 0$ is required for 
$|q| = e^{- \pi \mathrm{Im} \tau}<1$, 
which is a necessary condition for the definition of $\theta$ functions.
}
\begin{eqnarray}
& &\theta_1 (v, \tau) = q_0 q^{\frac{1}{4}} \frac{\mathbf{z} - \mathbf{z}^{-1}}{i} 
\prod_{k = 1}^{\infty} (1 - q^{2k} \mathbf{z}^2) (1 - q^{2k} \mathbf{z}^{-2}), \\
& & q_0 = \prod_{k=1}^{\infty} (1 - q^{2k}), \quad 
q = e^{ i \pi \tau}, \quad \mathbf{z} = e^{i \pi v}, \quad \mathrm{Im}
 \tau > 0.
\end{eqnarray}
Then the 1-lump solution on $T^2$ is given in the simple closed form: 
\begin{eqnarray}
u^{(1)}(z, \beta, \gamma) = 
\frac{\theta_1 (i \beta^{-1} (z - 2 \lambda), i |\beta^{-1}\gamma|)}
{\theta_1 (i \beta^{-1} (z - 2 \nu), i |\beta^{-1} \gamma|)},
\label{eq:T2-1lump}
\end{eqnarray}
where $\tau = i |\beta^{-1} \gamma|$ and 
$\mathrm{Im} \tau = |\beta^{-1} \gamma| > 0$.
The expression is valid even for the case $\beta^{-1}\gamma < 0$.
Again, we do not need any regularization for the multiplication of the
solution \eqref{eq:S1-1lump}.
Using the property of the theta function, 
\begin{eqnarray}
& & \theta_1 (i \beta^{-1} z - 1, i \beta^{-1} \gamma) = 
- \theta_1 (i \beta^{-1} z, i \beta^{-1} \gamma), \\
& & \theta_1 (i \beta^{-1} z + i \beta^{-1} \gamma, i \beta^{-1} \gamma) 
= - e^{2 \pi \beta^{-1} z} e^{\pi \beta^{-1} \gamma} \theta_1 (i
\beta^{-1} z, i \beta^{-1} \gamma),
\end{eqnarray}
the periodicity of the solution \eqref{eq:T2-1lump} is found to be
\begin{eqnarray}
& & u^{(1)} (z + i \beta, \beta, \gamma) = u^{(1)} (z, \beta, \gamma), \\
& & u^{(1)} (z + \gamma, \beta, \gamma) = e^{- 4 \pi \beta^{-1} (\lambda - \nu)}
 u^{(1)}(z, \beta, \gamma).
\end{eqnarray}
Therefore, in general, the solution \eqref{eq:T2-1lump} is not periodic
in the $\gamma$ direction. Only the twisted boundary condition is
allowed in that direction when the parameters satisfy $\mathrm{Re}
(\lambda - \nu) = 0$.
For the solution \eqref{eq:T2-1lump}, we find 
\begin{eqnarray}
\partial \log W^{\dagger} W 
&=& - i \beta^{-1} 
\frac{\theta_1 (i \beta^{-1} (z - 2 \lambda))}{\theta_1 (- i
\beta^{-1} (z - 2 \nu))} 
\times 
\nonumber \\
& & 
\frac{
\theta'_1 (- i \beta^{-1} (z - 2 \lambda)) 
\theta_1 (- i \beta^{-1} (z - 2 \nu))
-
\theta_1 (- i \beta^{-1} (z - 2 \lambda)) 
\theta'_1 (- i \beta^{-1} (z - 2 \nu))
}
{
|\theta_1 (i \beta^{-1} (z - 2 \nu))|^2
+
|\theta_1 (i \beta^{-1} (z - 2 \lambda))|^2
},
\nonumber 
\end{eqnarray}
where the theta functions have a common 
modulus $\tau = |\beta^{-1} \gamma |$.
Since the function $\theta_1 (v,\tau)$ has a zero at $v = 0$ and no
poles in the defined region (the fundamental lattice $- \frac{\sqrt{2}}{2} \gamma \le x \le 
  \frac{\sqrt{2}}{2} \gamma, - \frac{\sqrt{2}}{2} \beta \le y \le 
  \frac{\sqrt{2}}{2} \beta$), the pole of the function $U$ is at $z = 2
  \nu$.
The residue at the pole is evaluated as $\mathrm{Res}_{z = 2 \nu} \partial
\log W^{\dagger} W = 1$, which implies $Q=1$.

Now, let us consider the decompactification limits of the solution 
in the real and imaginary directions.
Using the expansion of the $\theta$-function,
\begin{eqnarray}
\theta_1 (v, \tau) = 2 q^{\frac{1}{4}} q_0 \sin \pi v 
\prod_{n=1}^{\infty} (1 - 2 q^{2n} \cos 2\pi v + q^{4n}),
\end{eqnarray}
and the fact, $q = e^{- \pi \beta^{-1} \gamma} \to 0$ in the limit 
$\gamma \to \infty$, we find 
\begin{eqnarray}
\lim_{\gamma \to \infty} 
u^{(1)} 
(z, \beta, \gamma) 
= \frac{\sin i \pi \beta^{-1} (z - 2 \lambda)}{\sin i \pi \beta^{-1} (z - 2 \nu)} 
= \frac{\sinh \pi \beta^{-1} (z - 2 \lambda)}{\sinh \pi \beta^{-1} (z - 2 \nu)} 
= u^{(1)}  (z, \beta).
\end{eqnarray}
This is just the array of the 1-lump solution on $\mathbb{R}^2$ along
the imaginary direction. 
Next, using the Jacobi identity relation of the $\theta$-functions, 
the decompactification limit along the real direction is calculated as 
\begin{eqnarray}
\lim_{\beta \to \infty} 
u^{(1)}  (z, \beta, \gamma) &=& 
\lim_{\beta \to \infty}
\frac{e^{\pi \beta^{-1} \gamma^{-1} (z-2\lambda)^2}}
{e^{\pi \beta^{-1} \gamma^{-1} (z-2\nu)^2}} 
u^{(1)}  (- i \gamma^{-1} \beta z, i \gamma^{-1} \beta)
\nonumber \\
&=& 
\lim_{\beta \to \infty}
\frac{e^{\pi \beta^{-1} \gamma^{-1} (z-2\lambda)^2}}
{e^{\pi \beta^{-1} \gamma^{-1} (z-2\nu)^2}}
\frac{
2 \tilde{q}^{\frac{1}{4}} \tilde{q}_0 \sin \pi \gamma^{-1} (z - 2
\lambda) 
\prod_{n=1}^{\infty} 
(1 - 2 \tilde{q}^{2n} \cos 2 \pi \gamma^{-1} (z - 2 \lambda) + \tilde{q}^{4n})
}
{
2 \tilde{q}^{\frac{1}{4}} \tilde{q}_0 \sin \pi \gamma^{-1} (z - 2
\nu) 
\prod_{n=1}^{\infty} 
(1 - 2 \tilde{q}^{2n} \cos 2 \pi \gamma^{-1} (z - 2 \nu) + \tilde{q}^{4n})
}
\nonumber \\
&=& 
\frac{\sinh \pi (i\gamma)^{-1} (z - 2 \lambda)}{\sinh \pi (i\gamma)^{-1} (z - 2
\nu)}, 
\end{eqnarray}
where we have defined 
$\tilde{q} = e^{2 \pi i (-1/\tau)} = e^{- 2 \pi \gamma^{-1} \beta}$.
The result is the 1-lump solution aligned along the real direction
with interval $\gamma$ as expected.

Generalization to the multilump solutions is straightforward.
The $n$-lump solution on $T^2$ is given by 
\begin{eqnarray}
u^{(n)} (z, \beta, \gamma) = 
\prod_{k=1}^n
\frac{\theta_1 (i \beta^{-1} (z - 2 \lambda_k), i |\beta^{-1}\gamma|)}
{\theta_1 (i \beta^{-1} (z - 2 \nu_k), i |\beta^{-1} \gamma|)}.
\label{eq:T2-nlump}
\end{eqnarray}
Its periodicity is 
\begin{eqnarray}
& & u^{(n)} (z + i \beta, \beta, \gamma) = u^{(n)} (z, \beta, \gamma), \\
& & u^{(n)} (z + \gamma, \beta, \gamma) = e^{- 4 \pi \beta^{-1} \sum_{k=1}^n (\lambda_k - \nu_k)}
 u^{(n)}(z, \beta, \gamma).
\label{torus_periodicity}
\end{eqnarray}
Therefore, when the following condition 
\begin{equation}
\sum_{k=1}^n \lambda_k = \sum_{k=1}^n \nu_k, \quad \lambda_k \not=
 \nu_k,
\label{eq:periodicity}
\end{equation}
is satisfied, the solution becomes periodic. This is possible only for
the $n \ge 2$ cases, thus confirming the mathematical result on 
harmonic maps.
Note that when one relaxes the condition \eqref{eq:periodicity} and the
parameters satisfy $\mathrm{Re} \sum_{k=1}^n (\lambda_k -
\nu_k) = 0$, the exponential factor in \eqref{torus_periodicity} 
becomes a phase factor and the solution satisfies the twisted boundary condition.

When the periodicity condition 
\eqref{eq:periodicity}
is satisfied, we expect that the
solutions can be rewritten as elliptic functions. 
For example, when we choose $\nu_1 = \nu_2 \equiv \nu$ and $\lambda_1 =
\nu - \frac{i \gamma}{4}, \lambda_2 = \nu + \frac{i \gamma}{4}$ 
for $n=2$ case, the solution \eqref{eq:T2-nlump} is rewritten as 
\begin{eqnarray}
& & u^{(2)} (z, \beta, \gamma) 
= - 4 \beta^{2} 
\left(
\frac{\theta_4^0}{\theta_1^{0 \prime}}
\right)^2 
\left\{
\wp (2 (z - 2 \nu)) - e_1
\right\}, 
\nonumber \\
& & e_1 = 
\left(
\frac{\pi}{2 \beta}
\right)^2 \frac{1}{3} 
\left(
(\theta_2^0)^2 + (\theta_3^0)^2
\right), \quad \theta_l^0 \equiv \theta_l (0, \tau), \ (l=1,2,3), \nonumber \\
& & \tau = | \beta^{-1} \gamma |, 
\end{eqnarray}
where $\wp$ is the Weierstrass $\wp$ function, the degree-2
elliptic function. The moduli space of this 2-lump solution on
$T^2$ was studied in \cite{Sp}. 

\begin{figure}[t]
 \begin{center}
 \includegraphics[scale=0.8]{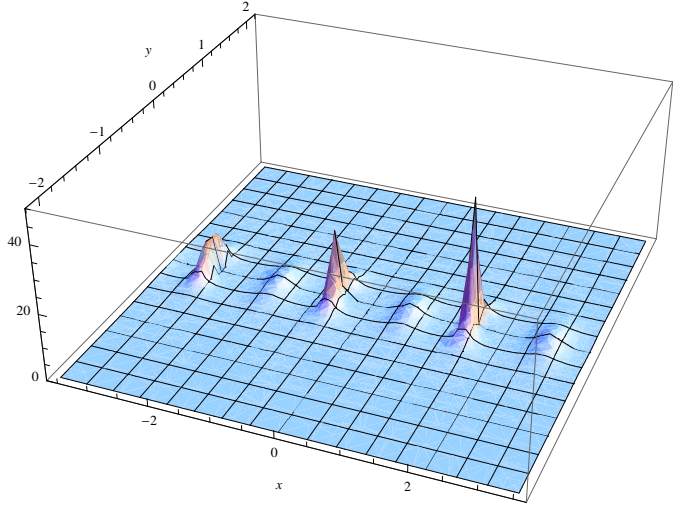} 
 \end{center}
 \caption{
Energy density for the solution \eqref{eq:T2-nlump} in the one
  fundamental lattice $- \frac{\sqrt{2}}{2} \gamma \le x \le 
  \frac{\sqrt{2}}{2} \gamma, -\frac{\sqrt{2}}{2} \beta \le y \le \frac{\sqrt{2}}{2} \beta$.
$\beta = 3, \gamma = 5$. $\lambda_1 = - 0.253, \lambda_2 = - 1.19,
  \lambda_3 = - 0.680, \nu_1 = - 0.918, \nu_2 = 0.629, \nu_3 = -0.680.$ 
The parameters $\lambda_i, \nu_i$ are a numerical solution to the
  modular invariance constraints \eqref{eq:modular_condition}.
}
 \label{fig:T2-lump}
 \end{figure}

The profile of the energy density for the $n=3$ solution is found in fig
\ref{fig:T2-lump}.
There are ``interference fringes'' among the three peaks since the lumps
are trapped on the finite-size lattice and the notion of the
``well separated'' is essentially lost in fully compact spaces.

Next, we study the modular invariance of the multilump solutions on a torus.
Let us consider a torus endowed with a generic modulus $\tau$.
A torus $T^2_{\tau}$ with modulus $\tau \in \mathbb{C}$ is defined by the 
equivalence class $ z \sim z - \beta (m + n \tau)$, $\beta \in \mathbb{R}$, $m, n
\in \mathbb{Z}$. The torus is invariant under the $PSL(2, \mathbb{Z})$
modular transformation,
\begin{eqnarray}
\tau \to \tau' = \frac{a \tau + b}{c
 \tau + d}, \quad 
ad - bc = 1, \quad a,b,c,d \in \mathbb{Z}.
\label{eq:modular}
\end{eqnarray}
Following the same procedure as before, the 
$n$-lump solution in the torus $T^2_{\tau}$ is found to be
\begin{equation}
u^{(n)} (z, \tau) = \prod_{k=1}^n 
\frac{\theta_1 (\beta^{-1} (z - 2 \lambda_k), \tau)}
{\theta_1 (\beta^{-1} (z - 2 \nu_k), \tau)}.
\label{eq:n-lump_T2}
\end{equation}
The modular transformation \eqref{eq:modular} is generated 
by the following fundamental transformations,
\begin{equation}
\tau \to \tau + 1, \quad 
\tau \to - \frac{1}{\tau} \ \textrm{with} \ z \to \tau z.
\end{equation}
Under the first transformation in the above, the $\theta$ function changes
as 
\begin{equation}
\theta_1 (v, \tau + 1) = e^{\pi i/4} \theta_1 (v, \tau).
\label{eq:modular1}
\end{equation}
The solution \eqref{eq:n-lump_T2} is therefore invariant
 under the transformation \eqref{eq:modular1}, cancelling
the phase factor $e^{\frac{\pi i}{4}}$.
Next, using the relation, 
\begin{equation}
\theta_1 (v, -1/\tau) = 
e^{i \pi v^2 \tau} e^{- 3\pi i/4} \tau^{1/2}
 \theta_1 (\tau v, \tau),
\end{equation}
the solution \eqref{eq:n-lump_T2} transforms as
\begin{equation}
u^{(n)} (z, \tau) \to u^{(n)} (z, - 1/\tau) 
= \exp 
\left[
4 \pi i \tau \beta^{-2}
\left\{
 (\sum_{i=1}^n \lambda_i - \sum_{i=1}^n \nu_n) z 
- (\sum_{i=1}^n \lambda^2_i - \sum_{i=1}^n \nu^2_i)
\right\}
\right]
u^{(n)} (\tau z, \tau).
\end{equation}
Therefore, the solution is 
modular invariant if the following conditions are satisfied:
\begin{equation}
\sum_{i=1}^n \lambda_i = \sum_{i=1}^n \nu_i, \quad 
\sum_{i=1}^n \lambda^2_i = \sum_{i=1}^n \nu^2_i, \quad 
\lambda_i \not= \nu_j \ \textrm{for all} \ i,j.
\label{eq:modular_condition}
\end{equation}
Again, $n=1$ is the special case.
It is easy to find that the condition \eqref{eq:modular_condition}
cannot be satisfied for the $n=1$ case.
When $n=2$, we find that the first two conditions in 
\eqref{eq:modular_condition} imply 
$\lambda_1 =
\nu_2, \lambda_2 = \nu_1$, which contradicts the third condition.
Therefore the modular invariance is
generically lost. On the other hand, there are infinitely many solutions
to the conditions \eqref{eq:modular_condition} for $n \ge 3$.
It is apparent that the modular invariance conditions \eqref{eq:modular_condition}
contain the periodicity condition \eqref{eq:periodicity}.
Hence the modular invariance is sufficient for the periodicity of the solutions.

Once the $n$-lump solutions satisfy the modular invariance conditions
\eqref{eq:modular_condition}, the solutions 
are generically rewritten as elliptic functions. 
To show this fact, 
let us consider the following relations between the
$\theta$ function and the Weierstrass $\sigma$ function:
\begin{eqnarray}
& & \sigma (2 \omega_1 z) = 2 \omega_1 e^{2 \eta_1 \omega_1 z^2} \theta_1
 (z, \tau)/\theta^{\prime 0}_1, 
\nonumber \\
& & \tau = \omega_3/\omega_1, \quad 
\eta_1 = \zeta (\omega_1) = \frac{\pi^2}{\omega_1} 
\left(
\frac{1}{12} - 2 \sum_{k=1}^{\infty} \frac{k q^{2k}}{1 - q^{2k}}
\right), \quad 
q = e^{\pi i \tau},
\end{eqnarray}
where $2 \omega_1, 2 \omega_3$ are two distinct periods of doubly periodic functions.
Then the $n$-lump solution \eqref{eq:T2-nlump} can be rewritten as 
\begin{equation}
u^{(n)} (z, \tau) = e^{A(z)}
\prod_{k=1}^n
\frac{
\sigma (2 \omega_1 \beta^{-1} (z - 2 \lambda_k))
}
{
\sigma (2 \omega_1 \beta^{-1} (z - 2 \nu_k))
},
\label{eq:T2-nlump_sigma}
\end{equation}
where the exponential factor is evaluated as 
\begin{eqnarray}
A (z)
&=& 8 \eta_1 \omega_1 \beta^{-2} 
\left[
\left(
\sum_{k=1}^n \lambda_k 
- 
\sum_{k=1}^n \nu_k 
\right) z
+
\left(
\sum_{k=1}^n \lambda_k^2
- 
\sum_{k=1}^n \nu_k^2
\right)
\right].
\end{eqnarray}
Applying the modular invariance conditions \eqref{eq:modular_condition},
this exponential factor vanishes and the solutions are totally expressed
by the elliptic functions.
The expression \eqref{eq:T2-nlump_sigma} 
is nothing but the solution discussed in \cite{CoZa}.
The contributions of these solutions to the partition function of the non-linear sigma
models on a torus are discussed in \cite{RiRo}. 
The Nahm transformation and moduli spaces of $\mathbb{C}P^{N-1}$ models
on a torus were discussed in \cite{AgAsWi}.
However, our solution \eqref{eq:T2-nlump} is more generic and constructive, allowing
the clear decompactification limits and the modular invariance.

So far we have focused on the base point condition $u^{(n)}(\infty) = 1$
on $\mathbb{R}^2$.
When we switch to the other base point conditions, for example $u^{(1)} (\infty) = 0$ on
$\mathbb{R}^2$, the solution on $T^2$ becomes
\begin{eqnarray}
u^{(1)} (z, \beta, \gamma) = i \eta (\tau) \theta_1^{-1} (i \beta^{-1} z,
 \tau),
\label{eq:reg_T2}
\end{eqnarray}
where we have again employed the $\zeta$ function regularization.
The function $\eta$ is the Dedekind $\eta$ function defined by 
\begin{eqnarray}
\eta (\tau) = q^{1/12} \prod^{\infty}_{k=1} (1 - q^{2k}), \quad q = e^{i
 \pi \tau}.
\end{eqnarray}
The periodicity of this solution is found to be
\begin{eqnarray}
& & u^{(1)} (z + i \beta, \beta, \gamma) = - u^{(1)} (z, \beta, \gamma), \\
& & u^{(1)} (z + \gamma, \beta, \gamma) = - e^{-2 \pi \beta^{-1} z} e^{- \pi
 \beta^{-1}} u^{(1)} (z, \beta, \gamma).
\end{eqnarray}
This solution does not show any modular invariance even for the $n \ge
3$ case. One can also find that the solution \eqref{eq:reg_T2} 
cannot be periodic even when the multilump generalization of the solution
\eqref{eq:reg_T2} is considered.
The properties of the solutions for different base point conditions 
are summarized in table \ref{tab:sol}.

\begin{table}[t]
\begin{center}
\begin{tabular}{|l||l|l|l|}
\hline
Base point cond. & Solution & Regularization & Modular inv. \\
\hline \hline
$u^{(n)} (\infty) = \infty$ & $\prod_{k=1}^n (i \eta (\tau))^{-1} \theta_1
	 (\beta^{-1} (z - 2 \lambda_k))$ & needed & lost \\
\hline
$u^{(n)} (\infty) = 1 $ & $\prod_{k=1}^n \theta_1 (\beta^{-1}
	 (z - 2 \lambda_k))/ \theta_1 (\beta^{-1}(z - 2 \nu_k))$ & no & exist \\
\hline
$u^{(n)} (\infty) = 0 $ & $\prod_{k=1}^n i \eta (\tau)/\theta_1
	 (\beta^{-1} (z - 2 \nu_k))$ & needed & lost \\
\hline
\end{tabular}
\end{center}
\caption{Solutions associated with each base point condition. The
 $\theta$ functions have common modulus $\tau$.}
\label{tab:sol}
\end{table}

Let us comment on the generalization of our construction 
to the $\mathbb{C}P^{N-1}$ models for $N \ge 3$ cases. 
One can easily find that this is straightforward.
The vector $W_i$ has $N-1$ independent components $w_{\hat{i}}$. 
Each component are holomorphic functions and we can construct solutions
on $T^2$ by the same way shown in the $N=2$ case.
The topological charges are determined by the highest degree of
the holomorphic functions $w_{\hat{i}} (z)$. 

\begin{figure}[t]
 \begin{center}
 \includegraphics[scale=0.7]{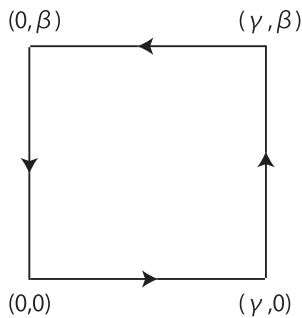} 
 \end{center}
 \caption{The closed path on the torus.}
 \label{fig:closed}
 \end{figure}

Finally, let us see the topological charge of the BPS lumps on a torus.
Without loss of generality, one can consider a rectangle torus 
defined by $z \sim z + (i \beta + \gamma)$.
The topological charge of lumps is given by the first Chern number
\begin{equation}
Q = - \frac{1}{4 \pi} \int \! d^2 x \ \varepsilon^{ab} F_{ab}.
\end{equation}
We demand that the $U(1)$ gauge field and, hence, the scalar field are periodic up to
the gauge transformation:
\begin{eqnarray}
\begin{aligned}
& A_1 (x_1, x_2 = \beta) = A_1 (x_1, x_2 = 0) 
- \partial_1 \lambda^{(2)} (x_1), \\
& A_2 (x_1 = \gamma, x_2) = A_2 (x_1 = 0, x_2) - \partial_2 \lambda^{(1)} (x_2).
\end{aligned}
\end{eqnarray}
Note that the gauge parameters $\lambda^{(m)} (x_n)$ depend only on $x_n
\ (n \not= m)$. Then the topological charge is given by
\begin{eqnarray}
Q = \frac{1}{2\pi} 
\left[
\lambda^{(1)} (\beta) - \lambda^{(1)} (0) + \lambda^{(2)} (0) -
\lambda^{(2)} (\gamma)
\right].
\end{eqnarray}
This is the gauge transformation parameter along the closed path
depicted in fig \ref{fig:closed}. 
On the other hand, once one goes around the closed path, the scalar
field acquires the phase $\lambda^{(1)} (\beta) - \lambda^{(1)} (0) + \lambda^{(2)} (0) -
\lambda^{(2)} (\gamma)$. 
The single-valuedness requires that 
this phase factor must be an integer multiple of
$2 \pi$. Therefore, the topological charge on $T^2$ must be integer,
\begin{equation}
Q = n, \quad n \in \mathbb{Z}.
\label{eq:top_charge}
\end{equation}
Configurations with nonzero topological number $Q$ are caused by the
large gauge transformation $\lambda^{(m)} (x_n)$ that is defined modulo $2\pi$.

\section{Conclusion and discussion}
In this paper, we have studied the topological BPS lumps in supersymmetric
$\mathbb{C}P^{N-1}$ non-linear sigma models on a torus $T^2$.
Following the philosophy of Harrington-Shepard, 
we have established the constructive procedure to give
the BPS lump solutions for arbitrary topological number
$Q=n$ by collecting the ``fundamental lumps'' aligned periodically.
The charge-$n$ BPS lump solutions on $T^2$ are obtained by
arranging the charge $n$ lumps on $\mathbb{R}^2$ at equal
intervals along two distinct directions. 
The function form of the solutions depends on the choice of the base point
condition of the 
fundamental lumps on $\mathbb{R}^2$.
Choosing the base point condition $u (\infty) = 0$ or $u (\infty) =
\infty$ requires the regularization of the infinite products of
rational maps. We have employed the $\zeta$-function regularization and found the
explicit solutions that exhibit suitable pole structures.

On the other hand, for the base point condition $u (\infty) = 1$, we 
do not need any regularization scheme. 
For the $n=1$ case, we have found that there is no solution that satisfies the
periodic boundary condition and the solution is not modular invariant anymore.
This is consistent with the statement that there is no
degree 1 elliptic functions on a torus. However, 
if the twisted boundary conditions are allowed, 
the solution turns out to be acceptable provided that the parameters of the solutions
are chosen appropriately. 
For $n=2$, there are no parameters $\lambda_i, \nu_i$ that satisfy the modular invariance conditions.
In the cases of $n \ge 3$, however, 
we find that there are infinitely many parameters that satisfy 
the modular invariance conditions.

Although the lumps on a torus were discussed in several contexts in the
past \cite{Su}, our construction is quite simple and constructive,
 and the solutions have the definite
decompactification limit by construction.
Since our construction of the solutions is so simple, 
we can obtain solutions on $T^2$, even for sigma models with other target
spaces, the same way.
Moreover, utilizing our construction, 
we expect that we can find solutions with nontrivial holonomy parameters
on compact spaces.
Such solutions on $\mathbb{R} \times S^1$ have been investigated
in \cite{Br}. 
When a solution has non-trivial holonomy 
along the compact
spaces, one expects that it has fractional topological charges.
This fact can be seen also in the gauge theory instantons in four dimensions.
It was discussed that gauge theories in a box (hypertorus) admit
instantons with a fractional Pontryagin number when the twisted boundary
conditions are imposed \cite{tH, Ba}.
These instantons have constituents in their inner parts.
For example, the constituents of doubly periodic instantons in $SU(2)$ Yang-Mills theories
are discussed in \cite{FoPa} and instantons with fractional charges are
studied in \cite{Mo}.

Finally, let us comment on the applications of our construction in the
other contexts. 
The two-dimensional supersymmetric sigma models are considered as 
 the effective action of a vortex in supersymmetric gauge
theories. Therefore the lumps on the compact spaces are interpreted as 
four-dimensional gauge theory instantons inside the vortex wrapping
the compact spaces. We will explore this possibility in the future works.
Time evolutions of the solutions on $T^2$ would be also 
interesting topics. 
Although, the generalization of our construction to $\mathbb{C}P^{N-1}$
with $N \ge 3$ is straightforward, its dynamics would
be different compared with the $N=2$ case as in the case of
$\mathbb{R}^2$ \cite{MaSu}.
The other time-dependent solutions, for
example, the Q-lumps \cite{Le2} on the torus can be constructed in the same
way.
Lumps with fractional topological charges and their constituents in gauged sigma models \cite{NiVi} 
and the other context \cite{EtFuGuKoNaNiOhVi} have been studied.
It would be interesting to investigate these kinds of fractional lumps in
the sigma models on a torus with twisted boundary conditions.

\end{document}